\RequirePackage[2020-02-02]{latexrelease}
\documentclass{article} 

\usepackage{preprint}

\usepackage{amsmath, amsthm, amssymb, amsfonts}
\newcommand{\norm}[1]{\left\lVert#1\right\rVert}
\usepackage{gensymb}
\usepackage[numbers,square,sort&compress]{natbib}
\usepackage[utf8]{inputenc}	
\usepackage[T1]{fontenc}	
\usepackage{xcolor}		
\usepackage[colorlinks = true,
            linkcolor = black,
            urlcolor  = blue,
            citecolor = cyan,
            anchorcolor = black]{hyperref}	
\usepackage{booktabs} 		
\usepackage{nicefrac}		
\usepackage{bm}
\usepackage{microtype}		
\usepackage{lineno}		
\usepackage{float}			

\usepackage{newfloat}
\DeclareFloatingEnvironment[name={Supplementary Figure}]{suppfigure}
\usepackage{sidecap}
\sidecaptionvpos{figure}{c}

\usepackage{titlesec}
\titlespacing\section{0pt}{12pt plus 3pt minus 3pt}{1pt plus 1pt minus 1pt}
\titlespacing\subsection{0pt}{10pt plus 3pt minus 3pt}{1pt plus 1pt minus 1pt}
\titlespacing\subsubsection{0pt}{8pt plus 3pt minus 3pt}{1pt plus 1pt minus 1pt}

\usepackage{tikz,xcolor,hyperref}

\definecolor{lime}{HTML}{A6CE39}
\DeclareRobustCommand{\orcidicon}{
	\begin{tikzpicture}
	\draw[lime, fill=lime] (0,0) 
	circle [radius=0.16] 
	node[white] {{\fontfamily{qag}\selectfont \tiny ID}};
	\draw[white, fill=white] (-0.0625,0.095) 
	circle [radius=0.007];
	\end{tikzpicture}
	\hspace{-2mm}
}
\foreach \x in {A, ..., Z}{\expandafter\xdef\csname orcid\x\endcsname{\noexpand\href{https://orcid.org/\csname orcidauthor\x\endcsname}
			{\noexpand\orcidicon}}
}

\usepackage{xwatermark}
\newwatermark[firstpage,color=gray!90,angle=0,scale=0.28, xpos=0in,ypos=-5in]{* Corresponding author: \href{mailto:depablo@uchicago.edu}{depablo@uchicago.edu} \\ $\dag$ These authors contributed equally to this work}

\title{LCPOM: Precise Reconstruction of \\ Polarized Optical Microscopy Images of Liquid Crystals}
\shorttitle{LCPOM}
\usepackage{authblk}

\author[1 $\dag$\orcidA{}]{Chuqiao Chen}
\author[1 $\dag$\orcidB{}]{Viviana Palacio-Betancur}
\author[2\orcidC{}]{Sepideh Norouzi}
\author[1\orcidD{}]{Pablo F. Zubieta Rico}
\author[2\orcidE{}]{Monirosadat Sadati}
\author[1,3,4\orcidF{}]{Stuart J. Rowan}
\author[1,4\orcidG{}]{Juan J. de~Pablo*}

\affil[1]{Pritzker School of Molecular Engineering, University of Chicago, Chicago, IL 60637}
\affil[2]{Department of Chemical Engineering, University of South Carolina, Columbia, SC 29208}
\affil[3]{Department of Chemistry, University of Chicago, Chicago, IL 60637}
\affil[4]{Materials Science Division, Argonne National Laboratory, Lemont, IL 60429}

\begin{document}

  
\maketitle

\begin{abstract}
When viewed with a cross-polarized optical microscope (POM), liquid crystals display interference colors and complex patterns that depend on the material's microscopic orientation. That orientation can be manipulated by application of external fields, which provides the basis for applications in optical display and sensing technologies. The color patterns themselves have a high information content. Traditionally, however, calculations of the optical appearance of liquid crystals have been performed by assuming that a single-wavelength light source is employed, and reported in a monochromatic scale. In this work, the original Jones matrix method is extended to calculate the colored images that arise when a liquid crystal is exposed to a multi-wavelength source. By accounting for the material properties, the visible light spectrum and the CIE color matching functions, we demonstrate that the proposed approach produces colored POM images that are in quantitative agreement with experimental data. Results are presented for a variety of systems, including radial, bipolar, and cholesteric droplets, where results of simulations are compared to experimental microscopy images. The effects of droplet size, topological defect structure, and droplet orientation are examined systematically. The technique introduced here generates images that can be directly compared to experiments, thereby facilitating machine learning efforts aimed at interpreting LC microscopy images, and paving the way for the inverse design of materials capable of producing specific internal microstructures in response to external stimuli.
\end{abstract}
\vspace{0.35cm}



\section{Introduction}\label{sec:intro}
When confined between a pair of linear polarizers, liquid crystals (LCs) can display a wide range of interference colors and complex patterns owing to the material's optical birefringence  (\textit{i.e.} the difference between refractive indices parallel and perpendicular to the molecular axis).~\cite{Andrienko2018,Gennes1993} The brightness and color hues are sensitive to the local molecular order, which can be controlled through external stimuli, including electric fields, magnetic fields, flows, chemical cues and temperature.~\cite{Andrienko2018,Gennes1993,HernandezOrtiz2011,Wu1991} On account of their responsive nature and large optical birefringence, LCs are widely used in optical devices, ranging from mature technologies such as liquid crystal displays to state-of-the-art sensors that can detect toxins, biomolecules, and microplastics.~\cite{Miller2013,Miller2013a,Chen2018,Carlton2013,Mukherjee2023,LopezLeon2011}

Liquid crystals are generally characterized using polarized optical microscopy (POM), which provides a direct measure of the material's alignment and and is able to identify any topological defects that might arise in a sample.\cite{Carlton2013, Shechter2020, MartinezGonzalez2015, Sadati2020, Lagerwall2012, You2019} In confined LC systems, large spatial distortions in the order field can develop on account of the incompatibility between surface and bulk orientations, leading to distinct POM images. The realignment of a confined LC can be triggered by altering the balance between elastic and surface energies; a minute change in the external environment can completely change the material's appearance under POM.\cite{Shechter2020, Tomar2012, Zhou2016} Sensing and display devices often rely solely on the transition between configurations that exhibit different topological defects (such as bipolar and radial), which are identifiable through the brightness profile. The substantial color changes that accompany such transitions are rarely exploited.\cite{You2019,Norouzi2022} Understanding how the POM color patterns of LCs correspond to a particular molecular alignment is of interest not only from a fundamental perspective, but also for development of next-generation display and sensing devices. 

One method to understand the color texture of POM images is to rely on the Michel-Levy chart, which tabulates the interference color as a function of thickness and birefringence.~\cite{Chen2020,Murphy2012,FernandezNieves2007} That method, however, is limited to a uniform orientation of the director field, and is incapable of predicting the interference color in confined geometries where the alignment exhibits large spatial variations. In addition, the POM images can change with light sources and viewing angles, making it difficult to match the color patterns with the underlying order field and hindering comparisons to experiments with different setups.

\begin{figure}[!htbp]
    \centering
    \includegraphics[width=0.98\textwidth]{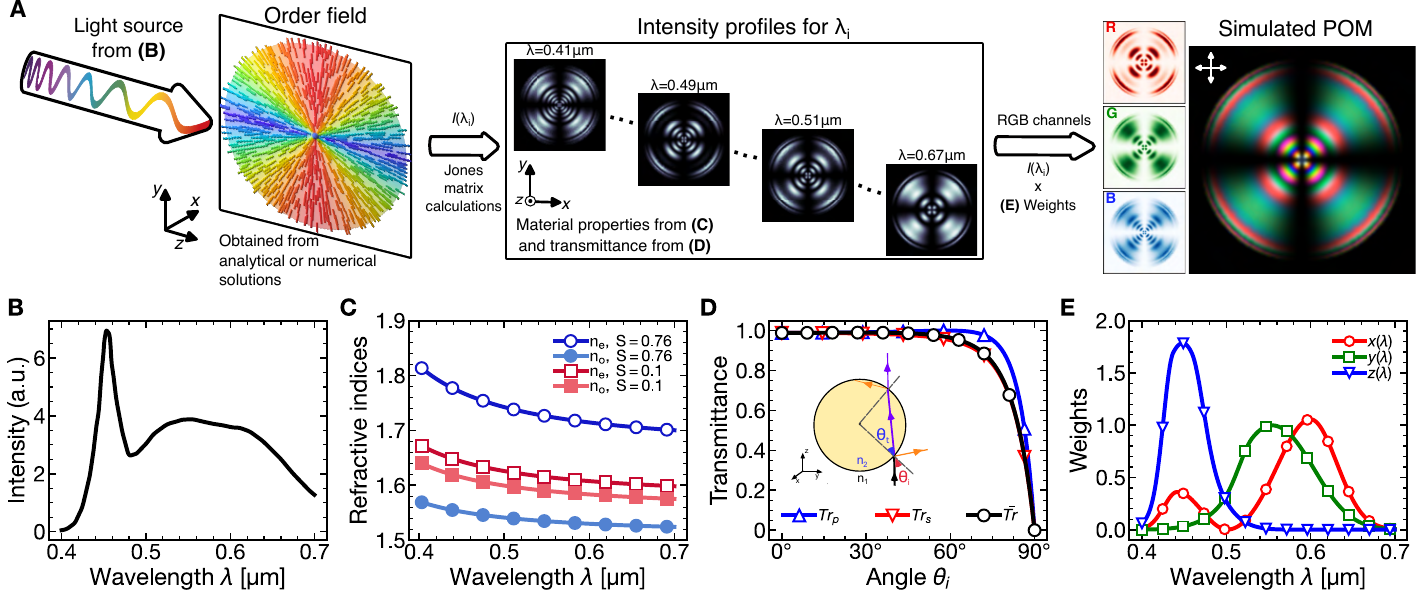}
    \caption{Illustration of the method and plots of key physical properties. (A) Schematic representation of the method for calculation of a colored POM image from the order field. In this illustration, a radial droplet with $d=20\mu m$ is computed with $N_{\lambda}=14$ wavelengths. (B) LED light spectrum obtained from the manufacturer. (C) Refractive indices of 5CB as a function of order parameter $S$ and wavelength $\lambda$. (D) Transmission ratio as a function of incident light angle. $Tr_p$, $Tr_s$ and $\overline{Tr}$ stand for transmission ratios of p-polarized, s-polarized light and the weighted average value. See Supplementary materials for details. (E) Color matching functions from the CIE 1931 standard.}
    \label{fig:fig1}
\end{figure}

More generally, POM images are calculated without color information using the Ondris-Crawford method, which produces the brightness profile corresponding to the LC order field.~\cite{Bellare1990,OndrisCrawford1991,Ellis2019} In this method, the sample is first discretized into layers whose thickness is much smaller than the wavelength. Subsequently, the propagation of light is modeled by multiplying the Jones matrix of each layer, which computes the retardation according to the local LC alignment. The method is versatile and easy to implement. It has been applied to many geometries, including droplets and toroids, where it is possible to reproduce the brightness profile of both nematic and cholesteric LCs.~\cite{Zhou2016,OndrisCrawford1991,Ellis2019,Jeong2014,Mur2017} The Ondris-Crawford method can be viewed as the standard approach for comparisons between experimental POM images and model predictions with numerical simulations.\cite{Shechter2020,Sadati2020,Zhou2016,Krakhalev2017,Krakhalev2019,Mur2016,Zumer2014} A major limitation, however, is that the original formulation assumes a single wavelength, and it is difficult to compare a monochromatic brightness profile with the color images that are typically obtained from a white light source with a distribution of wavelengths in the range between 400-680nm. Note that the effect of having a broadband light source has been discussed in several experimental and simulation studies, but reports that include simulated colored POM images have been limited and the agreement with experiments has been limited .~\cite{Mur2017,Krakhalev2017,Krakhalev2019,Mur2016,Geng2016} An exception is provided by the work of Yoshioka \textit{et al.}, who presented several colored POM images from calculations that showed good agreement with experiments.~\cite{Yoshioka2018} Unfortunately, few details regarding the calculation of the color images were provided in that report. In this work, we present a systematic methodological study of the computational generation of POM images (and an accompanying software package), which is validated through quantitative comparisons to experimental data for a variety of systems.

\section{Calculating optical textures}\label{sec:method}
The colored POM images are computed by introducing a color matching function that combines the information corresponding to multiple wavelengths to produce RGB values equivalent to the colors perceived by the human eye.~\cite{Mollon1999,Fairman1997,Broadbent2004,Wyman2013} In addition, we take into consideration the emission spectrum of the light source, the dependency of refractive indices on wavelength, and the reflection at the droplet interface. The accuracy and the applicability of our method are demonstrated by comparing simulated and experimental POM images of radial, bipolar, and cholesteric droplets. 

When colored POM images are captured in the laboratory, the sample is typically illuminated with a white light that has a non-uniform spectrum distribution. The light spectrum differs between laboratories and can alter the color texture.~\cite{David2018} To produce an accurate color image, the LED spectrum $I(\lambda)$ for experimental images produced in this work is obtained from the manufacturer or measured by an optical spectrometer. (\autoref{fig:fig1}A and Supplementary Fig. S1) In the calculations presented below, the light spectrum (400 nm - 680 nm) is discretized into $N_{\nu}=20$ intervals and the intensity profile for each wavelength is computed using the Ondris-Crawford method.~\cite{OndrisCrawford1991,Ellis2019}

The order field configurations are obtained either from analytical or numerical solutions, which are then interpolated onto a regular grid with the desired resolution. Each of the single-wavelength intensity profiles depends on the local LC alignment and the optical birefringence ($\Delta n$). (\autoref{fig:fig1}A) It is important to note that $\Delta n$ is not a single constant, but a function of the wavelength ($\lambda$) and the order parameter $S$ (which is a function of temperature and spatial gradient). Overall, both the ordinary and extraordinary refractive indices ($n_o$ and $n_e$ respectively) decrease with wavelength and saturate in the near-IR region. The quantitative relationship can be described by a three-band model with constants fitted to experimental measurements.~\cite{Wu1991,Wu1993} (\autoref{fig:fig1}C, see Supplementary Materials for the relevant equations).

In confined droplets, the director field is distorted and the local order parameter $S$ is smaller near the topological defects, leading to a local drop in $\Delta n$. Here we assume that the dependence of $n_e$ and $n_o$ on $S$ caused by the spatial gradient is equivalent to the variation of $S$ caused by temperature. The following relationship is adapted from the Vuks equation:~\cite{Li2004}
$$ \Delta n = (n_e -n_o)S/S_r$$
As an example, for the liquid crystal 4-cyano-4’-phentylbiphenyl (5CB), the reference state is taken to be $S_r=0.76$ at $T=25.1 \mathrm{^{\circ}}$. The refractive indices $n_o$ and $n_e$ are plotted as a function of $\lambda$ for uniform alignment ($S=0.76$) and near a topological defect ($S=0.1$) (\autoref{fig:fig1}C). It is worth noting that at $S=0.76$, $\Delta n$ decreases from 0.067 to 0.051 as $\lambda$ increases from $0.40 \mu m$ to $0.70 \mu m$, which is significant enough to affect higher order interference colors; this effect has generally been ignored in previous reports. 

In addition to the interference taking place in the bulk of the liquid crystal, we also consider the light transmission ratio at the water-LC interface. For simplicity, diffraction and refraction at the interface are ignored. In this study, the transmission ratio $\overline{Tr}$ is approximated from the Fresnel equation using the refractive indices of water and 5CB.~\cite{Wu1993, Pedrotti1993} (Details are available in Supplementary materials) $\overline{Tr}$ decreases with increasing incident angle ($\theta_i$), leading to lower brightness near the periphery of the droplet (\autoref{fig:fig1}F).  

To combine these multiple intensity profiles at different wavelengths into an RGB image, we consider how humans perceive color and how color images are stored in modern electronic devices. Briefly, the human eye can sense different wavelengths of light mainly with three types of cone cells in the retina.~\cite{Mollon1999} These signals are processed by an intricate neural network to generate a perception of color in the brain. Modern-day electronics represent color images by assigning tri-stimuli values such as RGB, XYZ, or HSV (Hue Saturation Value) to each pixel, so that digital displays can allow the human eye to perceive colors that are relatively independent of the device or the lighting environment. The matching functions $x_i$ ($\lambda$) that transform wavelength signals to XYZ values were originally determined by the International Commission on Illumination (in 1931 - CIE 1931 color space), and they are still widely employed today (\autoref{fig:fig1}E).~\cite{Fairman1997,Broadbent2004} In this work, the intensity profiles $P(\mathbf{r},\lambda)$ at $N_{\lambda}$ wavelengths are weighed by the light intensity $I(\lambda)$ and the matching functions to obtain an XYZ color image which is then converted to the RGB color image by a linear transformation.~\cite{Wyman2013} 

The image before transformation is calculated by: 
\begin{align}
    X_i(\mathbf{r})&=\int P(\mathbf{r}, \lambda)\overline{Tr}^2(\mathbf{r}, \lambda)I(\lambda)x_i(\lambda)d\lambda \nonumber \\ &\approx \sum^{N_\lambda}_j P(\mathbf{r}, \lambda_j)\overline{Tr}^2(\mathbf{r}, \lambda_j)w_{ij}(\lambda_j)
\end{align}

where $w_{ij}=\int_{\lambda_j}^{\lambda_{j+1} } I(\lambda)x_i(\lambda)d\lambda$ is the weight to the $i$th color channel (X, Y, Z) independent of the director field. $P(\mathbf{r},\lambda_j)$ represents the single-wavelength intensity profiles obtained from the Ondris-Crawford method.

In summary, the color image is obtained in four steps: 1) Generate the director field through analytical or numerical solutions and interpolate it onto a regular grid with the desired resolution. 2) Compute $N_{\lambda}$ intensity profiles for discretized wavelengths using the Ondris-Crawford method and multiply by the transmission ratio according to the local curvature. 3) Weigh by the LED spectrum and color matching functions to get the XYZ channel images. 4) Transform the image from XYZ color space to RGB color space and represent the result as a color image.

\section{Methodology and implementation}\label{sec:pypackage}

Each image ($110\times 110$ pixels) takes less than 20 minutes to compute on a single Intel Core i5 CPU processor. The sensitivity and speed of this method can serve for generating a data set that facilitates inverse design or machine learning.~\cite{Walters2019,Doi2019}

\section{Case studies}\label{sec:results}
In this section, we present the diverse applications of the LCPOM Python package, demonstrating its exceptional capabilities and providing benchmarks for computational tools seeking to align with experimental observations. Not only does our software serve as a convincing proof of concept, but it also offers invaluable insights into crucial considerations such as system size (\autoref{subsec:radial_size}), treatment of topological defects (\autoref{subsec:ring_defect}), probing the orientation of nematic morphologies (see \autoref{subsec:bipolar_droplet}), and moreover, the faithful reproduction of POM images in cholesteric systems (see \autoref{subsec:cholesteric}).

\subsection{Effect of system size}\label{subsec:radial_size}
To explore the effect of system size, we have generated POM images of radial droplets with various diameters ($d$). The radial droplet provides a canonical example of the interplay between bulk elasticity and surface orientation. In the presence of a surfactant, the LC molecules orient perpendicular to the surface of the droplet, resulting in a hedgehog defect at the center of the droplet. This particular system has been reported to be stable across a wide range of temperature, different materials, and across different length scales. The analytical description of the radial director field is $\mathbf{n}=\mathbf{x}/\norm{\mathbf{x}}$. The defect is a divergence of the director field, as pictured in \autoref{fig:fig1}A.

\begin{figure}[!htbp]
    \centering
    \includegraphics[width=0.98\textwidth]{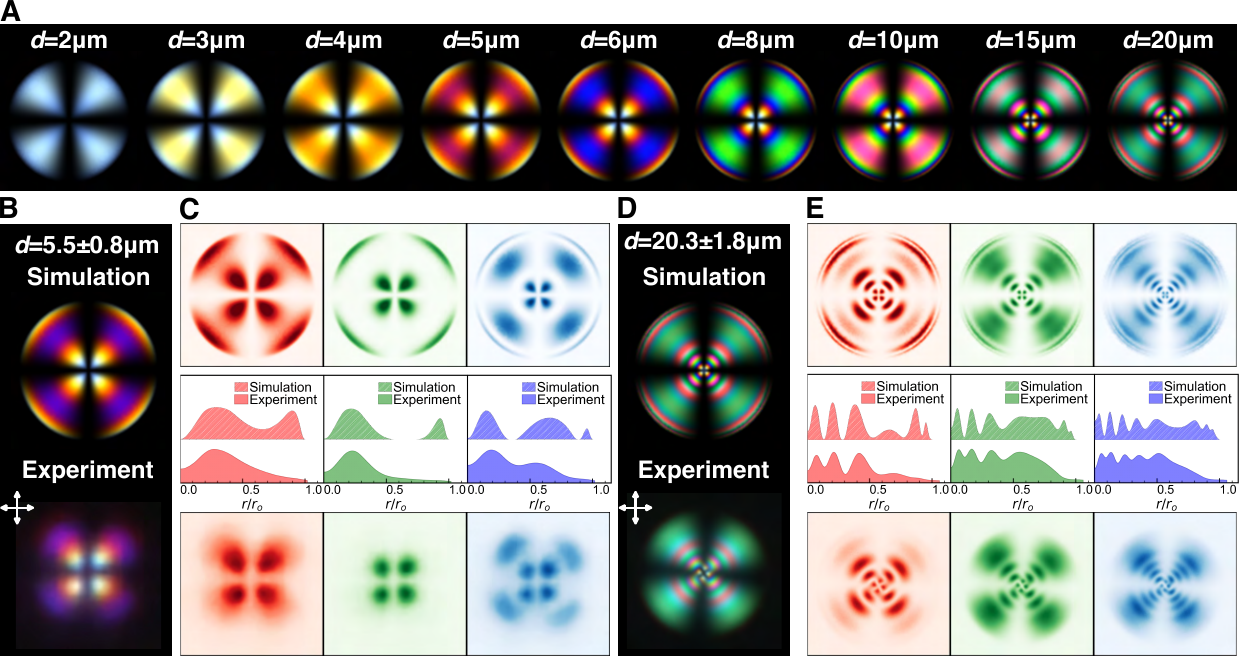}
    \caption{Simulation of the color POM images of radial droplets from the analytical order field. (A) Simulated images of a radial droplet from different sizes. (B) Comparison of the image of $d=5.5 \pm 0.8 \mu$m between simulation and experiment. (C) Intensity profiles of the RGB color channels and the radial intensity profile in B. (D) Comparison of the image of $d=20.3 \pm 1.8 \pm \mu$m between simulation and experiment. (E) Intensity profiles of the RGB color channels and the radial intensity profile in D. }
    \label{fig:fig2}
\end{figure}

The LCPOM results in \autoref{fig:fig2}A show the progression of colors that emerge as the diameter of the LC droplets is increased from $d=2\mu$m to $20\mu$m. As the diameter increases from $d=3\mu$m to $6\mu$m, the first-order interference colors change from yellow to blue (Fig. S2). At $d=6.5\mu$m, a second color ring emerges from the center of the droplet due to the spatial variation in the optical path differences. The simulated color textures are highly sensitive to system size, and a discrepancy in diameter as small as $0.5\mu m$ leads to notable differences in the optical appearance. This level of sensitivity is not seen in BW intensity profiles, indicating that more information about the LC order is captured by our proposed method. (Fig. S3) Moreover, this result highlights the use of a computational tool related to color to determine the size of an experimental system, or the possibility to infer the order parameter from microscopy images. More simulated POM images of this case are provided in Fig. S2.

To reproduce the color texture of experiments in simulations, the birefringence must be tuned down by $5\%$. This difference is consistent with the fact that frustrated alignment in curved geometries can lead to an attenuation of the optical birefringence compared to the uniform bulk samples in which the refraction indices are measured. Note that the droplet has a diffuse boundary on the bright-field images, and the determination of size has an average error of $14\%$ calculated from the FWHM (full width at half maximum) of the boundary (see Fig. S5). This diffuse appearance is caused by the diffraction at the interface between materials with mismatching refractive indices.~\cite{Pedrotti1993}

Overall, the color dependence on size in simulated images is in good agreement with experimental results (\autoref{fig:fig2}B, D and S4). The simulated and experimental images were decomposed into their respective RGB channel contributions (\autoref{fig:fig2}C and E). Given the symmetry of the radial structure, a quantitative comparison between simulated and experimental images can be achieved by performing a polar transformation and extracting the radial intensity profiles. Each of the RGB channels contains contributions from the entire light spectrum and does not have a simple analytical form. Importantly, the peak positions of the RGB color channels agree with the ones observed in experiments for $r/r_0<0.8$, which leads to a precise prediction of the color ring locations, even for large droplets, which exhibit higher-order interference. On the other hand, peaks close to the boundary ($r/r_0>0.8$) are not observed in experiments, as the intensity decays faster towards the edge of the droplet than predicted. This is attributed to the diffraction and fluctuation at the interface which, as discussed above, are not considered in our algorithm.

\begin{figure}[htbp]
    \centering
    \includegraphics[width=0.98\textwidth]{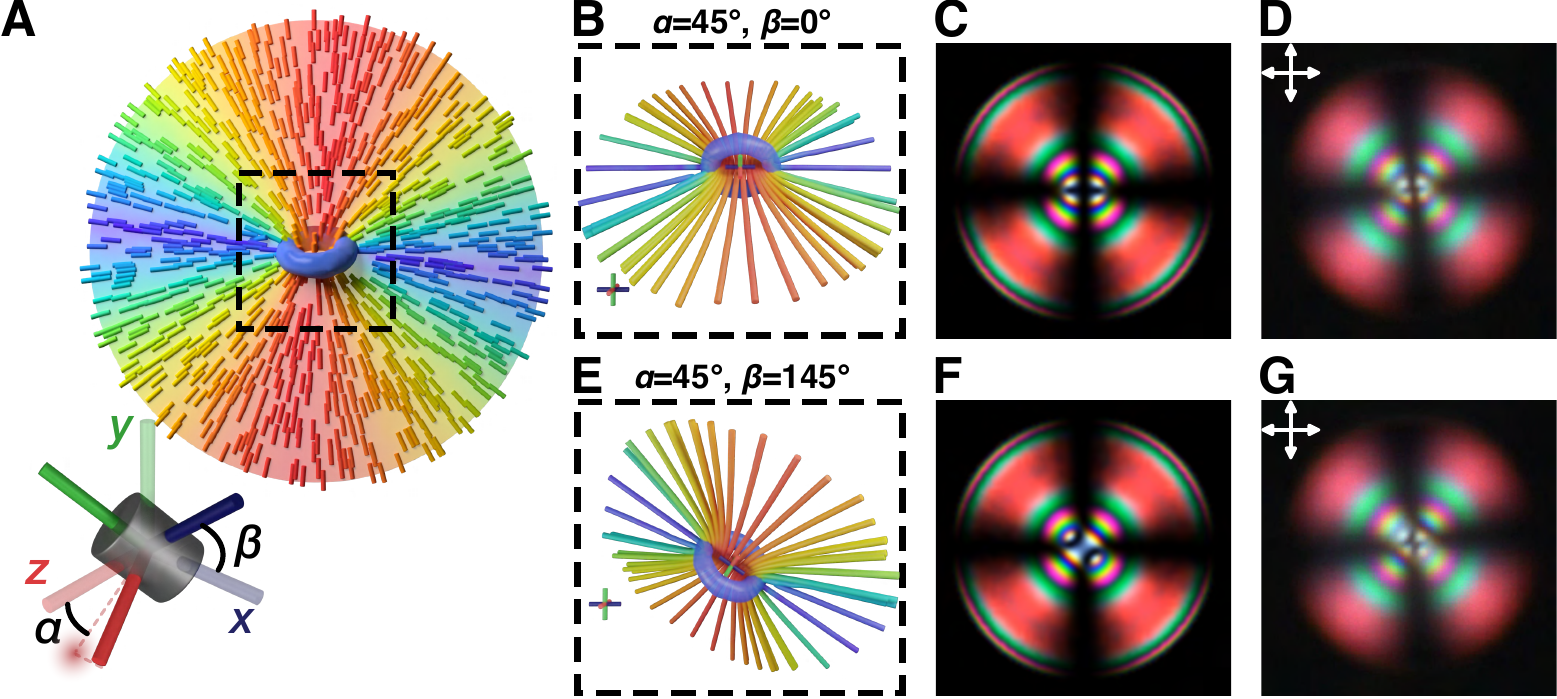}
    \caption{Effect of loop orientation in a radial droplet. (A) Simulated radial droplet with a disclination loop (regions with $S<0.3$) at the center. The diameter of the loop is approximately $1.9 \mu$m. (B-C) Schematic representation of  the orientation of the defect loop with $\alpha=45\degree$ and $\beta=0\degree$ and the corresponding simulated POM image with $d=17.5 \mu$m. (E-F) Schematic representation of the orientation of the defect loop with $\alpha=45\degree$  and $\beta=145\degree$  and the corresponding simulated POM image with $d=17.5 \mu$m.  (D) and (G): Experimental images for $d=17.5 \pm 1.6 \mu$m. }
    \label{fig:fig3}
\end{figure}

\subsection{Effects of non-point defects} \label{subsec:ring_defect}
The nature of the defect core is an active field of study, with implications for the assembly of colloids, active liquid crystals, and photonics. By definition, the defect core is the divergence of the vector order field, and can often be treated as a point charge. However, it has been shown theoretically and in experiments that the topological defects usually do not take the shape of a point, but appear as a region with diminished order parameters.~\cite{Tomar2012,Norouzi2022,ArmasPerez2015,Hernandez2012,ArmasPerez2015a,Mkaddem2000}. Radial droplets of 5CB typically exhibit loop disclinations whose diameter is sensitive to the anchoring strength, the elastic constants (temperature), and the size of the droplet.~\cite{Mkaddem2000,Terentjev1995,Kleman2006} The exact topology of the defect and the source of fluctuations have been long-standing questions that have attracted considerable theoretical interest.~\cite{Mkaddem2000,Terentjev1995,Kleman2006,Volovik1983,Lavrentovich1987} In experiments, the defect is usually small and sometimes appears as a blurry dot due to limitations in the optical resolution. In contrast to the hedgehog defect from \autoref{subsec:radial_size}, a loop disclination is surrounded by a continuous variation of the director field and has lower rotational symmetry. This implies that the optical texture should reflect when a rigid body rotation of the disclination loop occurs, as previously proposed by de la Cote \textit{et al.}~\cite{Cotte2022}

To examine this hypothesis, we performed simulations of nematic droplets under homeotropic anchoring conditions. Calculations of the order field following a Ginzburg-Landau relaxation yield a scalar and vector order field. The simulation details can be found in the Supplementary Information. The orientations in 3D can be described by two angles, $\alpha$ and $\beta$, because the order fields obey the $D_{\infty h}$ symmetry. Here, $\alpha$ is the out-of-plane tilt angle between the symmetry axis and the $xy$-plane and $\beta$ represents the in-plane rotation angle between the $0$-projection of the symmetry axis and the $y$ axis (\autoref{fig:fig3}A). 

As the droplet rotates, the reorientation of the loop creates subtle but clear changes in the POM image. As expected, the image bears the highest symmetry at $\alpha=90\degree$ and demonstrates more fuzzy central patterns compared to the POM image of the analytical form (Fig. S6). A distortion in the optical texture is observed when $\alpha$ deviates from $90\degree$ (\autoref{fig:fig3}C-F, Supplementary Movie 1). Importantly, the simulations produce color patterns that are very similar to those observed in experiments for particular orientations of the droplet (\autoref{fig:fig3}B-G, Supplementary Movie 2). This agreement suggests that under appropriate conditions the experiments can be directly compared to simulations to infer the orientation of the defects, thereby offering a new way of studying the dynamics and order fluctuations in confined LC environments.

\subsection{Elucidation of ambiguous micrographs through perspective sweep}\label{subsec:bipolar_droplet}
Another ubiquitous configuration in LC microemulsions is the bipolar droplet. It is characterized by two antipodal surface defects that emerge to satisfy a parallel molecular orientation tangential to the droplet's surface. The transition from a bipolar to a radial configuration can be triggered by adding a surfactant, which is the principle of operation for many LC-based sensing devices.~\cite{Shechter2020}  Similar to the order field in \autoref{subsec:ring_defect}, we performed numerical simulations to generate the vector and scalar order fields of a bipolar droplet that were then used as input for LCPOM. An advantage of this computational tool is the control over the viewpoint of an input morphology; different orientations that yield uncommon micrographs can be probed by this approach. In this case, simulated images were compared to experimental POM images of droplets created by dispersing 5CB in a PVA/water solution; additional experimental details are provided in the SI.                           

\begin{figure}[htbp]
    \centering
    \includegraphics[width=0.98\linewidth]{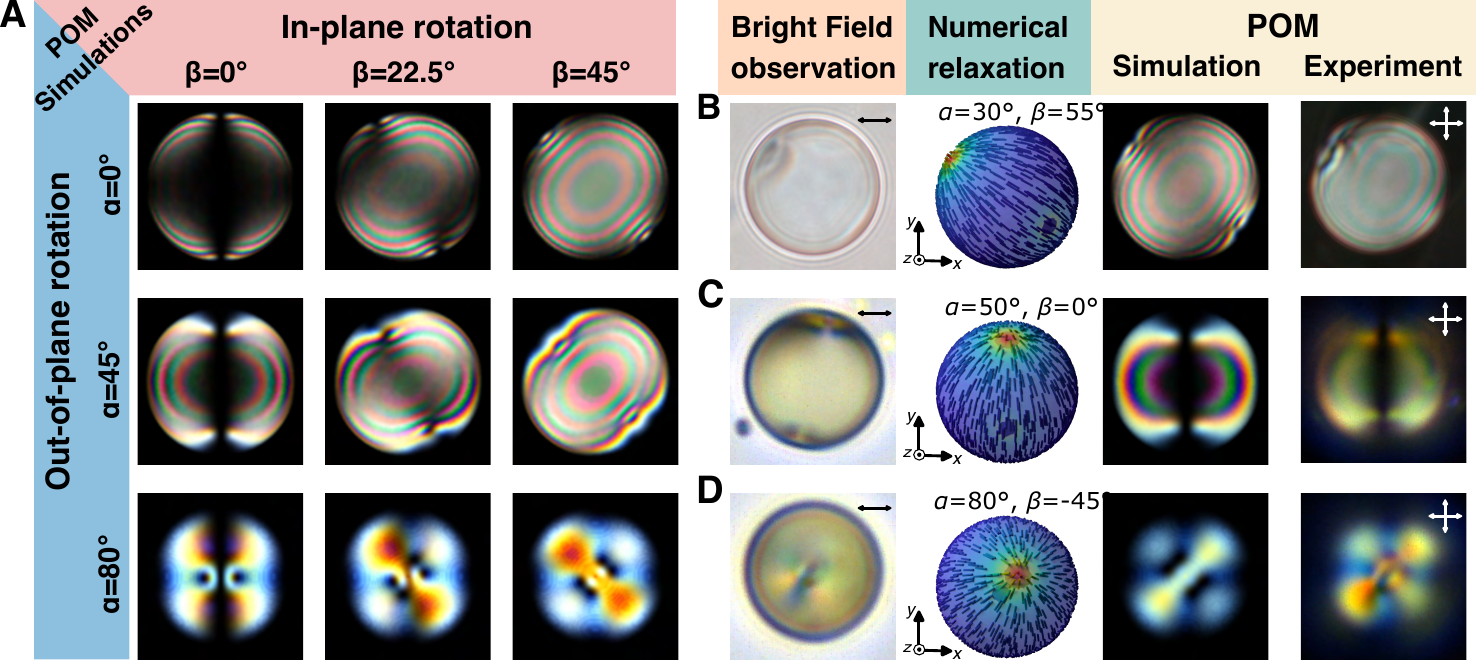}
    \caption{POM images of bipolar droplets with different configurations from simulations and experiments. (A) Effects of varying the orientation of a bipolar droplet with $d=20\mu$m. (B-D), from left to right: bright field image, order field from numerical relaxation, cross-polarized image, simulated POM image. (B) Droplet with a diameter of $22.9 \mu$m. The poles are oriented close to the $xy$-plane with a small out-of-plane tilt. The simulated image is produced with $\alpha=30\degree$  and $\beta=55\degree$. (C) Droplet with a diameter of $13.3 \mu$m with the bipolar axis coinciding with the $y$-axis. The simulated image is produced with $\alpha=50\degree$ and $\beta=0\degree$. (D) Droplet with a diameter of $12.4 \mu$m. The defects are located near the center of the $xy$-plane aligned with the $z$-axis. The simulated image is produced with $\alpha=80\degree$  and $\beta=-45\degree$. POM images in C and D are computed by using the light spectrum in Fig. S1(B).}
    \label{fig:fig4}
\end{figure}

The bipolar droplets have two defects at opposite poles that obey the $D_{\infty h}$ symmetry; all the orientations in 3D can be described by the $\alpha$ and $\beta$ angles from \autoref{subsec:ring_defect}. Simulated POM images for different orientations of the bipolar droplet are presented in \autoref{fig:fig4}A. In agreement with literature reports, simulated images at $\alpha=0\degree$ consist of concentric rings where the brightness and the outline change as the sample is rotated in the $xy$-plane. Optical textures similar to these are commonly reported in experiments.~\cite{Shechter2020,Volovik1983,Yang2022} Subtle deviations in the optical texture are often related to a small tilt of the bipolar axis, \textit{i.e.} the defects are tilted out of plane. For instance, it was found that the best agreement between simulations and experiments correspond to \autoref{fig:fig4}B, where $\alpha=30\degree$ and $\beta=55\degree$. Note that the spacing between the pink and green rings and the distortion features near the defects are all reproduced accurately. Interestingly, screening other orientations yields unfamiliar optical textures without the concentric rings. As such images have rarely been associated with bipolar configurations, we performed additional experiments to confirm the accuracy of LCPOM. Although textures with concentric rings are observed more often, morphologies corresponding to orientations with out-of-plane rotation $\alpha>45\degree$ (\autoref{fig:fig4}B-C) are also confirmed experimentally. It is possible that these textures are reported less frequently because they are difficult to classify. An alternative explanation is that bipolar droplets can adopt preferred orientations due to sedimentation or flow.~\cite{FernandezNieves2007} Nevertheless, the agreement between experiments and simulations suggests that LCPOM reliably generates POM images of bipolar droplets at arbitrary orientations, thereby providing a useful tool with which to interpret POM images and classify droplets that exhibit ambiguous optical morphologies. 

\subsection{LCPOM with a twist: cholesteric systems}\label{subsec:cholesteric}

\begin{figure}[htbp]
    \centering
    \includegraphics[width=0.98\textwidth]{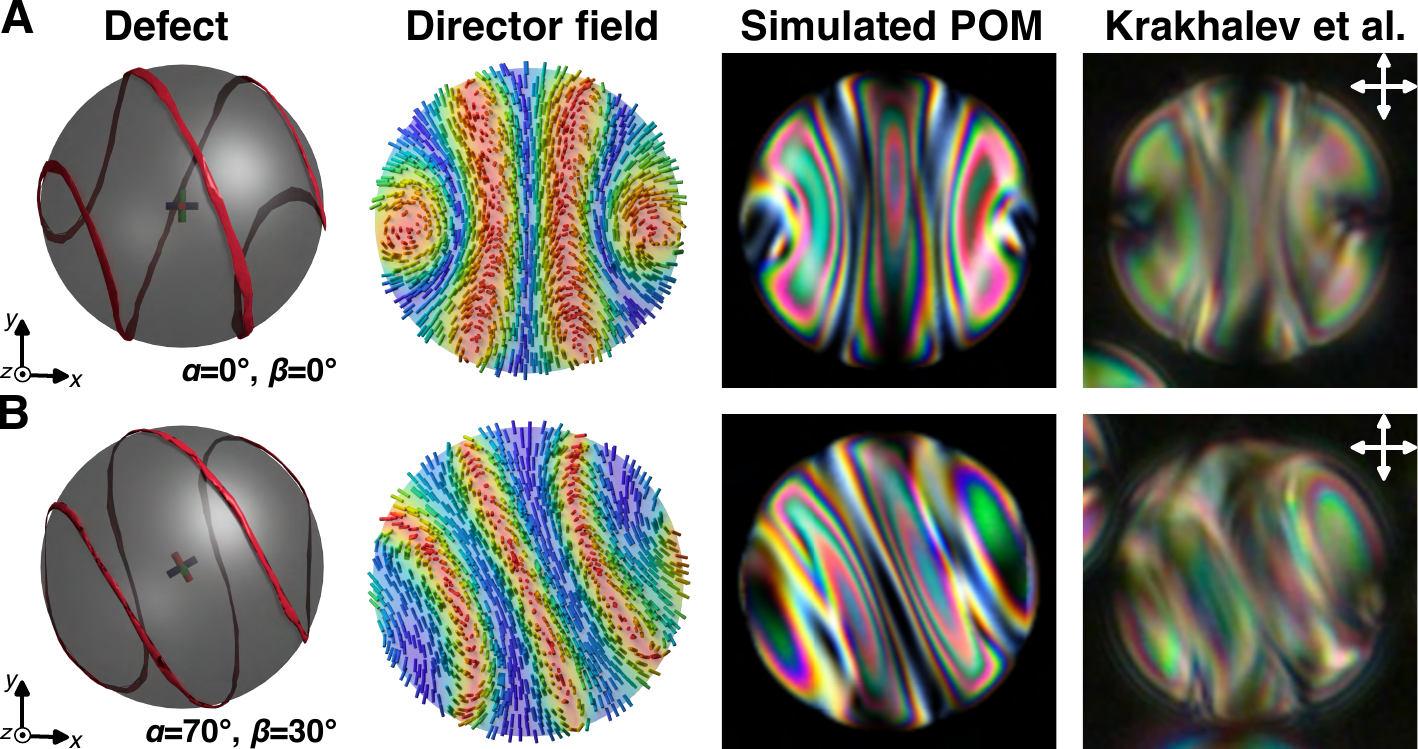}
    \caption{Comparison between simulated and experimental POM images of cholesteric droplets. The order fields of cholesteric droplets are obtained from Ginzburg-Landau relaxations with homeotropic boundary conditions. The number of turns is $N=3.8$ and the diameter of the droplet is set to the experimental value of $d=17 \mu m$. From left to right: isoclinical lines in red, order field, simulated POM images, and experimental images. The experimental images are adapted from Krakhalev \textit{et al.} Scientific Reports (2017).~\cite{Krakhalev2017} Order field in (B) is rotated by Euler angles $(\alpha,\beta)=(70\degree,30\degree)$ relative to (A). }
    \label{fig:fig5}
\end{figure}
Compared to nematics, cholesterics exhibit additional helical structures along the orientation of the director.~\cite{Zhou2016,Prishchepa2008} Depending on the droplet diameter, the helical pitch ($p_0$), and the surface interactions, cholesteric droplets can adopt complex internal morphologies and, as such, provide unique opportunities for engineering electro-optical and sensing devices.~\cite{Yoshioka2018,Lee2016,Xiang2016,PalacioBetancur2020} In general, the POM images of cholesteric droplets are highly sensitive to the droplet orientation, sometimes making it difficult to infer the exact underlying structure. Here, we computed POM images of a cholesteric droplet with a number of turns of $N=3.8$, and compared them to experimental images with $N=3.4$ reported by Krakhalev \textit{et al.}~\cite{Krakhalev2017}. The order field was obtained from a theoretically-informed Monte Carlo simulation with homeotropic boundary conditions, following the procedure of Palacio-Betancur \textit{et al.}~\cite{PalacioBetancur2020} The POM calculations are based on the material properties of the E7 mixture.~\cite{Krakhalev2017,Li2005} The images produced in this manner show excellent agreement with experiments.~\cite{Krakhalev2017,Krakhalev2019} Note that this match is only obtained when the birefringence is scaled down by $40\%$ compared to that of bulk E7, implying that the chiral dopant or the local twist may have led to a decrease in the birefringence. Another possible reason is that the droplets dispersed in polymer films may be oblate, and hence the optical path difference could be overestimated when a spherical geometry is assumed.~\cite{Krakhalev2017} To examine how the optical texture changes with droplet orientation, POM images were calculated at varying angles and compared with experimental results from Krahkhalev \textit{et al.} (Supplementary Movie 3).~\cite{Krakhalev2017} We found that a small change in the orientation can lead to distinctly different optical textures of the droplet, yet a good match with experimental results can be obtained when the orientation is set close to those reported in the original paper (\autoref{fig:fig5}D-F). In general, controlling the orientation of cholesteric droplets is challenging in experiments, making it difficult to investigate the optical texture systematically. By generating high-fidelity color images of complex structures, LCPOM can help develop a better understanding of structure in confined cholesterics, which is needed for design of advanced optical devices.

\section{Discussion and Conclusions}\label{sec:concl}
A straightforward method for simulation of color in POM images has been presented for confined liquid crystals, including droplets. By incorporating the emission spectrum of the light source, the dependence of refractive indices on wavelength, the transmission ratio at the droplet interface as well as the color matching functions, our simulation method is shown to be capable of generating colored POM images that are in quantitative agreement with experiments for radial, bipolar and cholesteric droplets. The method provides a particularly useful tool to validate theoretical models and to interpret experimental measurements. By comparing computed POM images of the order field profile obtained from theory or simulations to experimental POM images, one can gain insights into the governing physics and the balance of various phenomenological parameters. We envision that the proposed computational tool will help generate a realistic dataset to aid machine learning efforts aimed at understanding the structure and dynamics of liquid crystals, and will help engineer a new generation of LC-based sensing devices where color is used to extract detailed information about molecular-level sensing events.

\section*{Supplementary Information}
Experimental details, and further validation images are provided in the Supplementary Information.

\section*{Code and Data Availability}
The code for LCPOM along with its documentation will be released after beta testing. Sign up in \href{https://forms.gle/a56Hif8UzU4z9GFs5}{this form} to be notified once it is available.

The files containing the order fields and to reproduce the optical textures in this work are available upon request from the corresponding author.

\section*{Acknowledgments}
C.C. thanks Dr. Neil D. Dolinski for help on spectrum measurement. This work was primarily supported by the University of Chicago Materials Research Science and Engineering Center, which is funded by National Science Foundation under award number DMR-2011854. V.P.B. thanks the Fulbright commission in Colombia and COLCIENCIAS for support through the PhD student scholarship. M.S. is supported by National Science Foundation, Division of Materials Research, Condensed Matter Physics, under the NSF CAREER award 2146428. The authors also acknowledge the Research Computing Center of the University of Chicago for computational resources.


\normalsize
\bibliography{references}

\newpage
\appendix

\renewcommand\thefigure{S\arabic{figure}}   \renewcommand\thetable{S\arabic{table}}    
\setcounter{figure}{0}
\setcounter{table}{0}
\section{Materials and experimental methods}
To prepare nematic LC droplets, 4-cyano-4'-pentylbiphenyl (5CB, Sigma Aldrich) was rigorously vortexed in deionized water. Subsequently, 2wt\% of Poly (vinyl alcohol) (PVA, Mw=13k-23k, 87-89\% hydrolyzed, Sigma-Aldrich) and sodium dodecyl sulfate (SDS, Sigma-Aldrich) were added to the 5CB emulsion to stabilize the droplets and induce planar and homeotropic alignment, respectively. The droplets were then filled into rectangular borosilicate capillaries under capillary action ($0.1\times 0.2$ mm ID, VitroCom).

A cross-polarized optical microscope (Zeiss Axioscope 5, Oberkochen, Germany) equipped with an Axiocam 506 color camera and a 100X objective lens was employed to image the droplets in the trans- mission mode. The image intensity distribution for each channel of RGB was obtained using Zencore software. During the imaging process, the temperature of the droplets was controlled using a Linkam heating stage (model PE-120), which was connected to a temperature controller (model T-96) to maintain the temperature at 25\textdegree C.

Figures 4c-d were obtained in a different lab. To reduce the mismatch of refractive indices between 5CB and water, a water/glycerol mixture (50/50 vol\%) was used as the continuous phase to disperse the 5CB droplets. 2wt\% of Poly (vinyl alcohol) was added to induce planar anchoring. The droplets were filled into a circular cell (diameter = 8mm, depth = 0.2mm) to insulate flow. The images were obtained by Leica DM2700P polarized optical microscope equipped with MC170 HD 5 camera.

\section{Computational details}

\subsection{Refractive indices}
The extraordinary and ordinary refractive indices are computed from a three-band model that separates the contributions from $\sigma$ and $\pi$ electrons. The empirical constants $g_{0e}, g_{1e}, g_{2e}, g_{0o}, g_{1o}, g_{2o}, \lambda_0, \lambda_1, \lambda_2$ can be found in the original paper by Wu \textit{et al.} \cite{Wu1991,Li2005}

\begin{align}
n_e(\lambda) &= 1+ g_{0e}\frac{\lambda^2\lambda_0^2}{\lambda^2-\lambda_0^2}+ g_{1e}\frac{\lambda^2\lambda_1^2}{\lambda^2-\lambda_1^2}++ g_{2e}\frac{\lambda^2\lambda_2^2}{\lambda^2-\lambda_2^2} \\
n_o(\lambda) &= 1+ g_{0o}\frac{\lambda^2\lambda_0^2}{\lambda^2-\lambda_0^2}+ g_{1o}\frac{\lambda^2\lambda_1^2}{\lambda^2-\lambda_1^2}++ g_{2o}\frac{\lambda^2\lambda_2^2}{\lambda^2-\lambda_2^2}
 \end{align}

\subsection{Transmission ratio}
The transmission ratio depends on the state of polarization, the refractive indices ($n_1$ and $n_2$ for the continuous and dispersed phases respectively) and the incident angle $\theta_i$ of the light. In this study, the transmission ratio at the interface is approximated by the Fresnel equation.\cite{Pedrotti1993}
\begin{align} 
    Tr_s &= 1- \left| \frac{n_1\cos\theta_i - n_2\cos\theta_t}{n_1\cos\theta_i + n_2\cos\theta_t} \right|^2 \\
    Tr_p &= 1- \left| \frac{n_1\cos\theta_t - n_2\cos\theta_i}{n_1\cos\theta_t + n_2\cos\theta_i} \right|^2 \\
    \theta_t &= \sin^{-1}(n_1\sin\theta_i/n_2)
\end{align} 

Here, the refractive index of water is $n_1 = 1.33$. For simplicity, the refractive index of liquid crystal $n_2$ is taken to be an average value ($n_2 = (n_e + 2n_o)/3$). The Brewster angle is found to be around 59\textdegree, within 1\textdegree~ error from experimental values.\cite{Mur2016} The overall transmission $\overline{Tr}$ is given by:
\begin{equation}
\overline{Tr} (\theta_i, \lambda)=Tr_p \cos^2\theta_i+Tr_s \sin^2\theta_i
\end{equation}

\subsection{Continuum simulations of nematic liquid crystals}
We adopt a mean-field approach to simulating liquid crystalline order, described by the tensorial order parameter $\mathbf{Q}(\mathbf{x})$. The free energy functional includes a short-range Landau polynomial expansion in $\mathbf{Q}(\mathbf{x})$ to model the isotropic-nematic transition, a long-range functional that accounts for elastic distortions, and a surface contribution that imposes a preferred orientation at the boundaries.
\begin{align} \label{eq:free_energy}
F(\mathbf{Q} ) = \int d^3{\mathbf x}&\left[ \frac{A}{2}\left(1-\frac{U}{3}\right)\mathrm{tr}(\mathbf{Q}^2)-\frac{AU}{3}\mathrm{tr}(\mathbf{Q}^3)+\frac{AU}{4}\mathrm{tr}(\mathbf{Q}^2)^2 \right. \nonumber \\
&+\left. \frac{L_1}{2}\frac{\partial Q_{ij}}{\partial x_k}\frac{\partial Q_{ij}}{\partial x_k}+\frac{L_5}{2} \epsilon_{ikl} Q_{ij} \frac{\partial Q_{lj}}{\partial x_k} \right] \\ 
+ \oint & d^2{\mathbf x} \left[ \frac{W_{\parallel}}{2}\left(\mathbf{ \bar{Q}-\bar{Q}_{\parallel}}\right)^2+ \frac{W_{\perp}}{2}\left(\mathbf{Q-Q_\perp}\right)^2 \right]. \nonumber \end{align}

The model parameters include the Landau coefficient $A$ that sets the energy density scale, the dimensionless parameter $U$ that determines the IN transition (inverse temperature); $L_1$ and $L_5$ are the elastic constants related to splay and chiral twist deformations respectively, and $\epsilon_{ijk}$ is the Levi-Civita operator. The LC orientation at the surface is imposed to be homeotropic (perpendicular to the surface) or planar (tangential to the surface), and deviations from that orientation are penalized with magnitudes $W_{\perp}, W_{\parallel}$ respectively. The tensors $\bar{\mathbf{Q}}, \bar{\mathbf{Q}}_{\parallel}, \mathbf{Q}_{\perp}$ refer to projections of $\mathbf{Q}$ and are defined as follows: $\mathbf{Q}_{\perp}=S(\bm{\nu\nu-\bm{\delta}/3})$,  $\bar{\mathbf{Q}}_{\parallel}=\mathbf{p\cdot\bar{Q}\cdot p}$, $\bar{\mathbf{Q}}=\mathbf{Q}+S\bm{\delta}/3$, with $\mathbf{p}=\bm{\delta}-\bm{\nu\nu}$ and $\bm{\nu}$ is the normal unit vector on the surface. Further details of the free energy densities can be found in our previous works.~\cite{LondonoHurtado2015,ArmasPerez2015,ArmasPerez2015a,PalacioBetancur2020,PalacioBetancur2023} 

The free energy must satisfy the Euler-Lagrange equations, and the solution is found by allowing the tensor order parameter to evolve towards equilibrium, following a Ginzburg-Landau relaxation of $\mathbf{Q}$ of the form,
\begin{align}
\frac{\partial \mathbf{Q}}{\partial t} = - \frac{1}{\gamma}\left[ \frac{\delta F}{\delta\mathbf{Q}} \right]^\text{ST}&, \label{eq:GL} \\
\text{with B.C.s } \left[\frac{\delta F}{\delta\nabla\mathbf{Q}}\cdot\bm{\nu}\right]^{ST} = 0 \label{eq:BC}&
\end{align}
where $\gamma$ is the LC rotational diffusion coefficient, and $\left[\mathbf{A}\right]^{ST}=(\mathbf{A}+\mathbf{A}^T)/2 - \text{tr}(\mathbf{A})\bm{\delta}/3$ is the symmetric traceless operator. The Volterra derivatives are defined by,
\begin{equation}
\frac{\delta F}{\delta\mathbf{Q}} =  \frac{\partial F}{\partial\mathbf{Q}} - \frac{\partial}{\partial\mathbf{x}} \cdot \frac{\partial F}{\partial\mathbf{\nabla Q}}.
\label{eq:functional-def}
\end{equation} 

Numerical solutions to the Ginzburg-Landau relaxation are calculated with in-house code. The different geometries are discretized into tetrahedral elements with Cubit~\cite{Cubit}. We adopt the Galerkin method of weighted residuals to transform \eqref{eq:GL} and \eqref{eq:BC} into a system of linear algebraic equations. Open-source libraries like libMesh~\cite{Kirk2006} and SuperLU~\cite{li05} allow us to implement the numerical solution method. The scalar $S(\mathbf{x})$ and director $\mathbf{n}(\mathbf{x})$ order fields are calculated from the eigenvalues and eigenvectors of $\mathbf{Q}(\mathbf{x})$ once the systems have reached e, and are used as input parameters in LCPOM.

\section{Supplementary figures and tables}

Figures 3-5 of the main text use numerical solutions of $\mathbf{Q}$ to calculate the color POM images. The finite element meshes are composed of quadratic tetrahedral elements. The mesh has $3\times 10^5$ elements, and additional mesh refinement was performed for the cholesteric droplet case, in which the mesh has $5\times 10^5$ elements. The simulation parameters to obtain the order fields are as follows,
\begin{table}[H]
    \centering
    \caption{Simulation parameters for numerical solutions of the order field}
    \begin{tabular}{lccccc}
        \toprule
         {\bf System} & $A$ (J/m$^3$) & $U$ & $L$ (pN) & $W$ (J/m$^2$) & {\bf Droplet size} ($\mu$m) \\ \midrule
         Radial     & $1\times 10^5$& 4 & 6 & $\perp, 10^{-3}$     & 17.5\\ 
         Bipolar    & $1\times 10^5$& 4 & 6& $\parallel, 10^{-3}$ & 20.0 \\
         Cholesteric& $1\times 10^5$& 3.5 & 16 & $\perp, 10^{-4}$     & 17.0 \\
         \,$\hookrightarrow p_0 = 17.0 \mu$m \\
         \bottomrule
    \end{tabular}
    
    \label{tab:my_label}
\end{table}

\begin{figure}[H]
    \centering
    \includegraphics[width=0.9\linewidth]{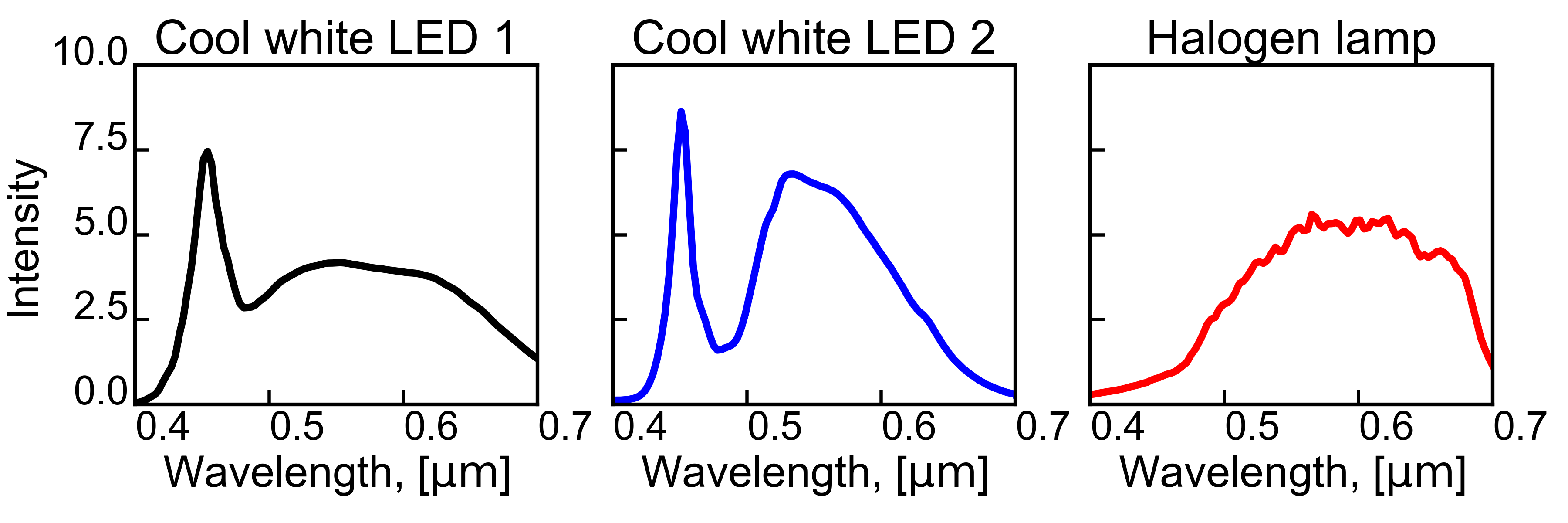}
    \caption{Microscope light sources}
    \label{fig:lightsource}
\end{figure}

\begin{figure}[H]
    \centering
    \includegraphics[width=0.95\linewidth]{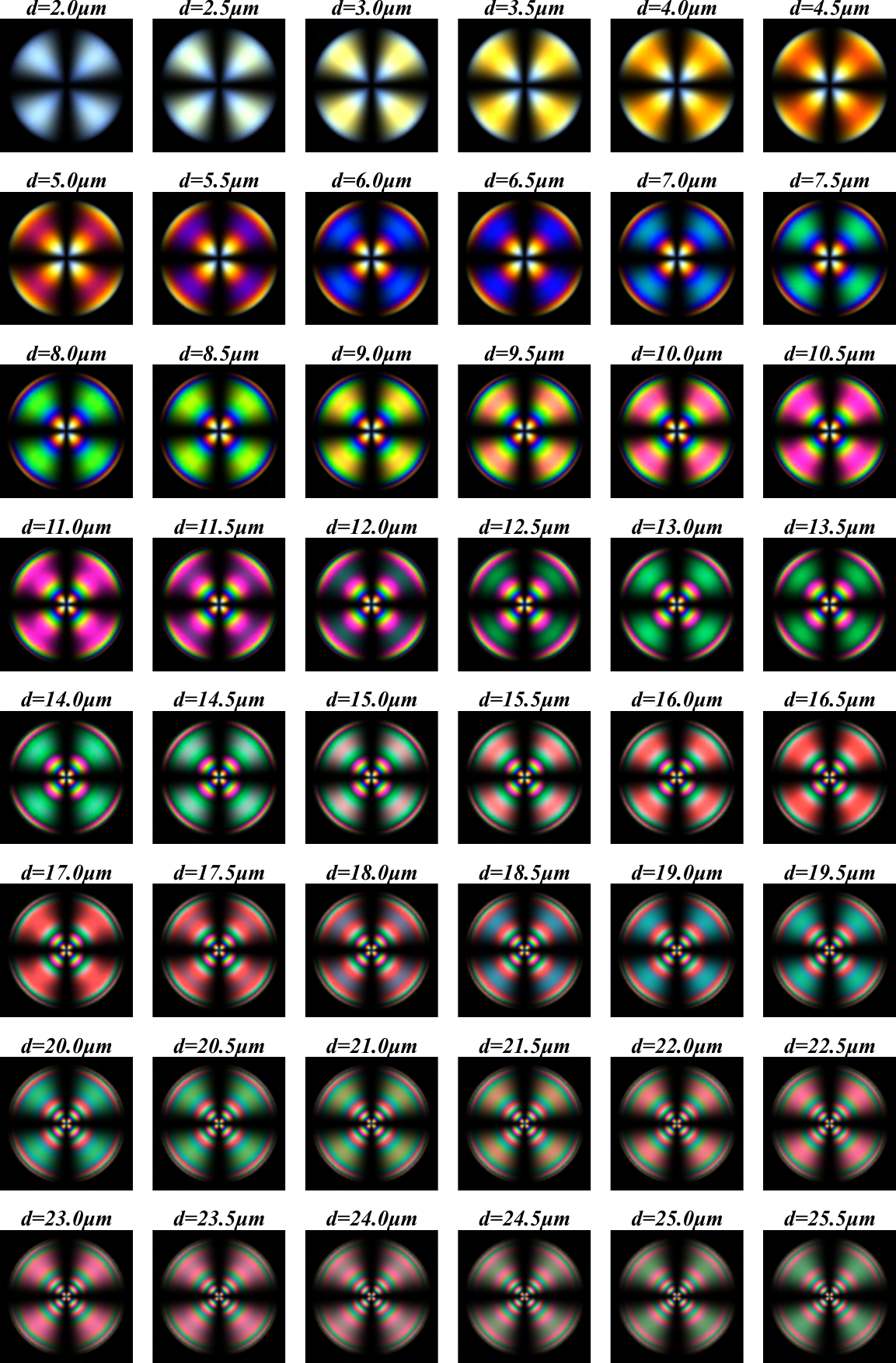}
    \caption{LCPOM simulations for radial droplets of various sizes}
    \label{fig:radial-sizes}
\end{figure}
\newpage
\begin{figure}[H]
    \centering
    \includegraphics[width=0.4\linewidth]{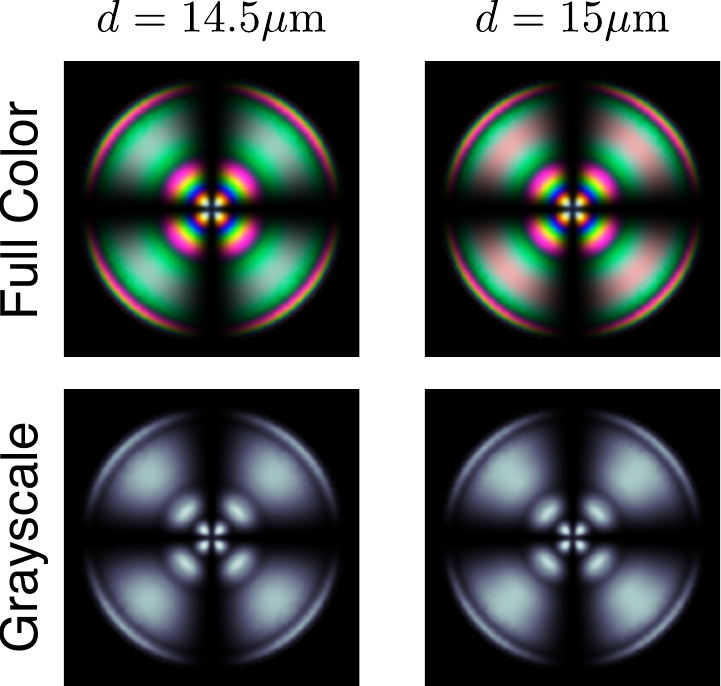}
    \caption{Comparison of sensitivity between color and BW POM images}
    \label{fig:color-bw}
\end{figure}

\begin{figure}[H]
    \centering
    \includegraphics[width=0.75\linewidth]{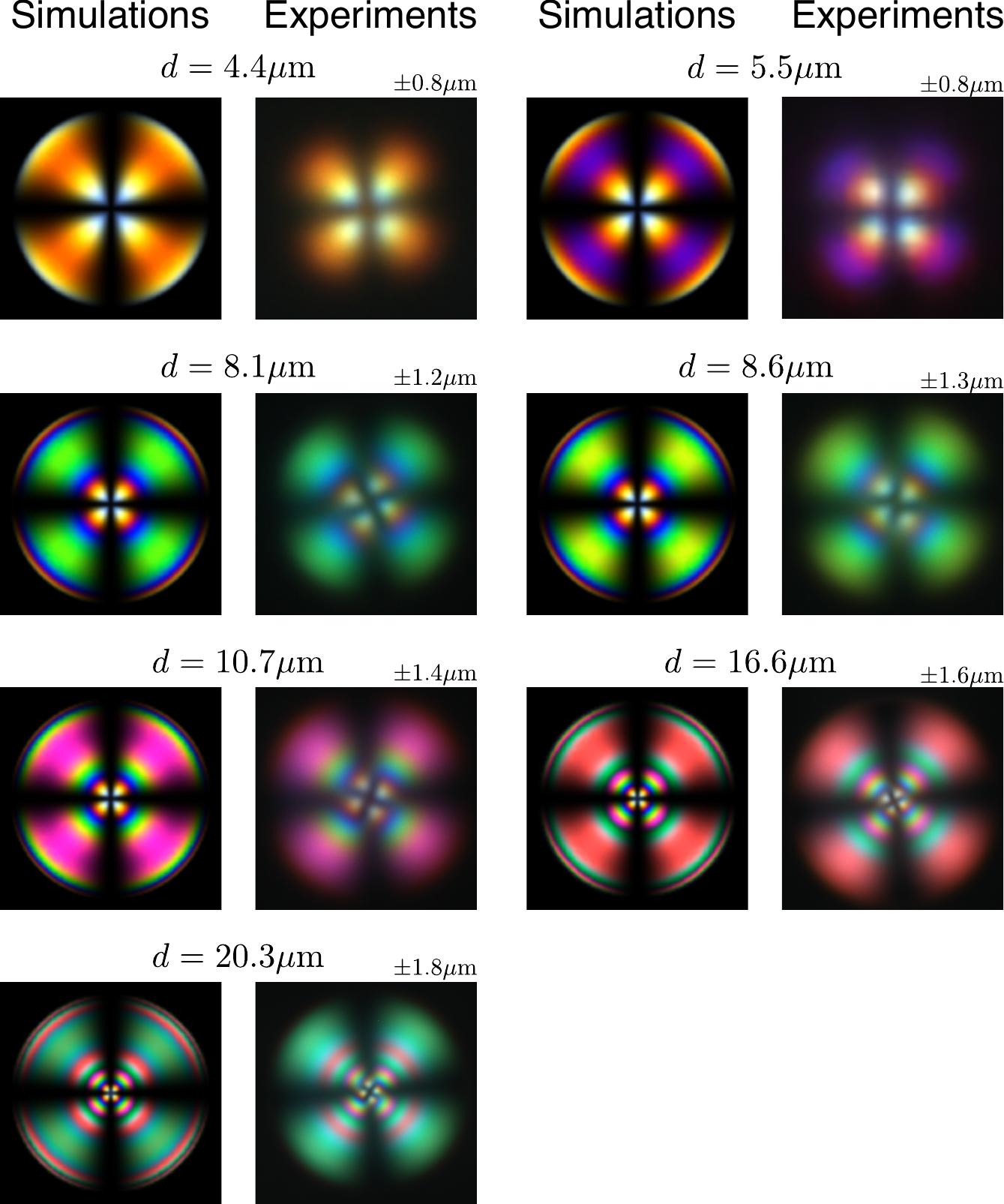}
    \caption{Comparison between experimental images of radial droplets and LCPOM simulations for the same size system}
    \label{fig:lcpom-exp}
\end{figure}

\begin{figure}[H]
    \centering
    \includegraphics[width=0.8\linewidth]{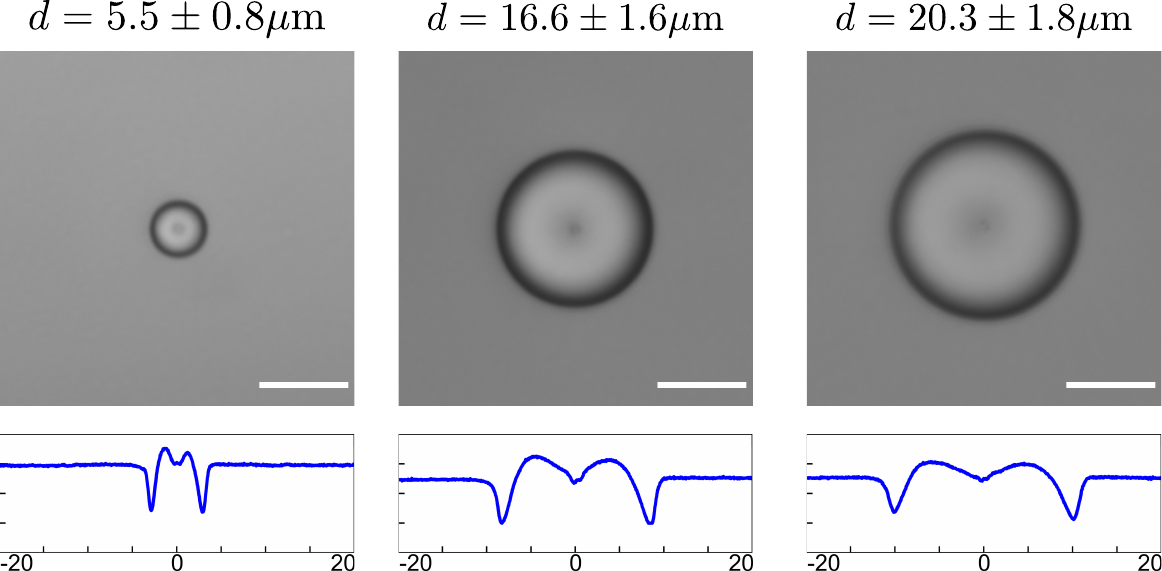}
    \caption{Size determination of radial droplets from bright field images}
    \label{fig:radial-size-BF}
\end{figure}

\begin{table}[H]
\centering
\caption{Size determination of radial droplets}
\begin{tabular}{ c c c c c c } 
  \toprule
   & Diameter $\mathrm{[\mu m]}$ & FWHM left $\mathrm{[\mu m]}$  & FWHM right $\mathrm{[\mu m]}$& FWHM mean $\mathrm{[\mu m]}$& \% error \\
  \midrule
1 & 4.4 & 0.6  & 0.9 & 0.8 & 17\% \\  
2 & 5.5 & 0.7  & 0.9 & 0.8 & 14\% \\ 
3 & 8.1 & 1.0  & 1.4 & 1.2 & 15\% \\ 
4 & 8.6 & 1.0  & 1.7 & 1.3 & 15\% \\ 
5 & 10.7 & 1.2  & 1.6 & 1.4 & 13\% \\ 
6 & 16.6 & 1.6  & 1.7 & 1.6 & 10\% \\ 
7 & 20.3 & 1.6  & 2.0 & 1.8 & 9\%\\ 
 \bottomrule
\end{tabular}
\end{table}

\begin{figure}[H]
    \centering
    \includegraphics[width=0.5\linewidth]{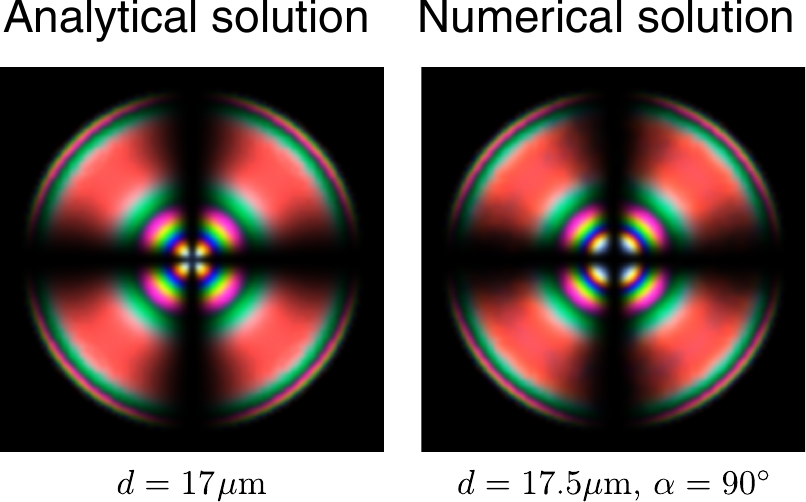}
    \caption[LCPOM simulations of analytical and numerical order fields]{LCPOM simulations of two radial droplets. Order fields obtained from analytical solution (hedgehog defect) and numerical solution (ring defect).}
    \label{fig:radial-ansatz-sim}
\end{figure}

\end{document}